\documentclass{aa}
\usepackage{graphicx}
\usepackage{txfonts,natbib}
\bibpunct{(}{)}{;}{a}{}{,}
\begin{document}
   \title{Long-term monitoring of the X-ray afterglow of GRB\,050408 with \textit{Swift}/XRT}


   \author{M. Capalbi\inst{1} \and D. Malesani\inst{2} \and M. Perri\inst{1}  \and  P. Giommi\inst{1,3} \and S. Covino\inst{4} \and  G. Cusumano\inst{5} \and V. Mangano\inst{5} \and T. Mineo\inst{5} \and  S. Campana\inst{4} \and G. Chincarini\inst{4,6} \and  V. La Parola\inst{5} \and A. Moretti\inst{4} \and P. Romano\inst{4} \and G. Tagliaferri\inst{4} \and L. Angelini\inst{7,8} \and P. Boyd\inst{7}\and D.N. Burrows\inst{9} \and O. Godet\inst{10} \and J.E. Hill\inst{7,11} \and J.A. Kennea\inst{9} \and F. Marshall\inst{7} \and P.T. O'Brien\inst{10} \and N. Gehrels\inst{7}}

   \offprints{M. Capalbi \\ \email{capalbi@asdc.asi.it}}

   \institute{ASI Science Data Center, via G. Galilei,
              I-00044 Frascati, Italy
	 \and International School for Advanced Studies (SISSA-ISAS),
              via Beirut 2-4, I-34014 Trieste, Italy
	 \and ASI, Unit\`a Osservazione dell'Universo, viale Liegi 26,
              I-00198 Roma, Italy
	 \and INAF -- Osservatorio Astronomico di Brera,
              via E. Bianchi 46, I-23807 Merate, Italy
	 \and INAF -- Istituto di Astrofisica Spaziale e Fisica Cosmica,
              Sezione di Palermo, via U. La Malfa 153, I-90146 Palermo, Italy
	 \and Universit\`a degli Studi di Milano-Bicocca, Dipartimento
              di Fisica, piazza delle Scienze 3, I-20126 Milano, Italy
	 \and NASA/Goddard Space Flight Center, Greenbelt, MD 20771, USA
	 \and Department of Physics and Astronomy, The Johns Hopkins University, 
              3400 North Charles Street, Baltimore, MD 21218, USA
	 \and Department of Astronomy and Astrophysics, Pennsylvania State
              University, University Park, PA 16802, USA
	 \and Department of Physics and Astronomy, University of Leicester,
              Leicester LE1 7RH, UK
         \and Universities Space Research Association, 10211 Wincopin Circle, Suite 500, 
              Columbia, MD, 21044-3432, USA
             }

   \authorrunning{M. Capalbi et al.}

  \titlerunning{ \textit{Swift}/XRT observations of the X-ray afterglow of GRB\,050408}
   \date{Received / Accepted}

   \abstract{

We present observations of the X-ray afterglow of GRB\,050408, a gamma-ray
burst discovered by HETE-II. \textit{Swift} began observing the field 42~min
after the burst, performing follow-up over a period of 38~d (thus spanning
three decades in time).
The X-ray light curve showed a steepening with time, similar to many other
afterglows. However, the steepening was unusually smooth, over the
duration of the XRT observation, with no clear break time.
The early decay was too flat to be described in terms of standard models. We
therefore explore alternative explanations, such as the presence of a structured
afterglow or of long-lasting energy injection into the fireball from the central GRB engine. 
The lack of a sharp break puts constraints on these two models. 
In the former case, it may indicate that the angular energy profile of the jet was 
not a simple power law, while in the second model it implies that injection did not 
stop abruptly. 
The late decay may be due either to a
standard afterglow (that is, with no energy injection), or to a jetted outflow
still being refreshed.
A significant amount of absorption was present in the X-ray spectrum,
corresponding to a rest-frame Hydrogen column density $N_{\rm H} =
1.2_{-0.3}^{+0.4} \times 10^{22}$~cm$^{-2}$, indicative of a dense environment.

\keywords{gamma rays: bursts -- X-rays: individual (GRB\,050408)} }

   \maketitle

\section{Introduction}
\label{intro}
\indent

The \textit{Swift} Gamma-Ray Burst Explorer \citep{gehrels},
designed to study gamma-ray bursts (GRBs), has unique characteristics allowing
the prompt observation of GRB afterglows
\citep[see][for a description of the \textit{Swift} instruments]{barthelmy,burrows05,roming}.
This has opened up the opportunity to
study the previously unexplored early phases of their evolution. 

On the other hand, being fully devoted to GRB studies, \textit{Swift} also has the
capability to perform detailed, long-term monitoring of afterglows. To
date, a number of GRBs have been observed up to $\sim 1$~month after the
explosion \citep[e.g. XRF\,050416A;][]{Mangano06}, allowing a systematic
study of the late phases of their evolution.
\textit{Swifts} rapid slewing capabilities also allows the prompt follow-up to
alerts coming from other satellites, such as HETE-II and INTEGRAL. GRB\,050408
was the first burst observed by \textit{Swift} which was not autonomously-triggered by the Burst Alert Telescope (BAT) {barthelmy} onboard.

%
GRB\,050408 was detected by HETE-II at 16:22:50.93 UT on 2005 April 8 and
localized by the Soft X-ray Camera (SXC) at 
RA(J2000)$\mbox{} = 12^{\mathrm h} 02^{\mathrm m} 15^{\mathrm s}$,
Dec(J2000)$\mbox{} = 10\degr 52\arcmin 01\arcsec$ \citep[80\arcsec{} error radius, 90\% containment;][]{Sakamoto05}.
The burst 
duration was $T_{90} \sim 34$~s in the 7--40 and 7--80 keV
bands, and $T_{90} \sim 15$~s in the 30--400 keV band. Spectral analysis of the prompt emission
showed that the fluence was $\sim 1.4 \times 10^{-6}$~erg~cm$^{-2}$ and $\sim 1.9 \times 10^{-6}$~erg~cm$^{-2}$
in the 2--30 keV and 30--400 keV energy bands, respectively.
Using the classification 
defined by \citet{Lamb05}, GRB\,050408 can therefore
be classified as an ``X-ray rich'' GRB \citep{Sakamoto05}.

On the basis of the HETE-II position, reported soon after the discovery via a
GCN notice, a target of opportunity (TOO) was uploaded to \textit{Swift} and the NFIs
were pointed at the target about 42~min after the trigger. The \textit{Swift} XRT
detected an uncatalogued, fading source inside the SXC error box
\citep{xrtgcn}, which was later refined with an error circle of 5\arcsec{} radius
\citep{xrtgcn2}.  Following the initial observation, XRT continued
monitoring the afterglow for several weeks, leading to a long, well-sampled
light curve.

Several 
follow-up observations were performed at other wavelengths. An
optical counterpart was discovered inside the SXC error circle
\citep{deugarte}, using the 1m and 6m telescopes at the Special Astrophysical
Observatory. This object was later seen to fade, confirming its afterglow
nature \citep{Huang05}. The astrometric position was provided by
\citet{Chen05}, who located the source at the coordinates RA(J2000)$\mbox{} =
12^{\mathrm h}02^{\mathrm m}17\fs328$, Dec(J2000)$\mbox{} =
+10\degr51\arcmin09\farcs47$ (0\farcs25 error radius).
A spectrum of the optical afterglow obtained with the LDSS-3 instrument
on the Magellan/Clay telescope showed the presence of emission and
absorption lines at a redshift $z = 1.236$ \citep{berger,berger_b}. The Gemini
telescope found a consistent value of $z = 1.2357 \pm 0.0002$
\citep{prochaska,Foley05}. Many other optical telescopes, including the
3.6m Telescopio Nazionale Galileo (TNG), detected and monitored the
optical afterglow. The analysis of the TNG data, taken simultaneously
with the XRT observation, is presented in a separate paper
\citep{tng}. Some of the afterglow properties have been discussed by
\citet{Foley05}.
At the location of the afterglow, UVOT detected a faint,
low-significance source in the coadded $U$-band image
\citep{holland}. No detection was possible in the other optical and
ultraviolet filters. Very Large Array radio observations at 8.5~GHz
revealed no sources at the afterglow position, down to a 2-$\sigma$
upper limit of 74~$\mu$Jy on 2005 Apr 9.26 UT \citep{radio}.

In the following sections we report a detailed analysis of the XRT
follow-up observations. In Sect.~\ref{xrt} we describe the XRT data
reduction and analysis, and in Sect.~\ref{discussion} we discuss the
results. A summary of our work is reported in
Sect.~\ref{conclusions}. Throughout the work, we follow the notation
$F_\nu(t,\nu) \propto t^{-\alpha}\nu^{-\beta}$ for the time and spectral
dependency of the flux. The times quoted are with respect  to the HETE-II
trigger time.

\begin{table}
\caption{X-ray afterglow light curve. The first column reports the center of the time bin, expressed in seconds since trigger.\label{tb:lcdata}}
\centering
\begin{tabular}{rrr}\hline\hline
\multicolumn{1}{c}{ Time }    & \multicolumn{1}{c}{Bin size} & \multicolumn{1}{c}{Count rate}       \\ 
\multicolumn{1}{c}{(s)}   & \multicolumn{1}{c}{(s)}          & \multicolumn{1}{c}{(10$^{-2}$ count~s$^{-1}$)} \\ \hline
  2612    &   120     &  23$ \pm $5	    \\
  2732    &   120     &  30$ \pm $5	    \\
  2852    &   120     &  29$ \pm $5	    \\
  6302    &   500     &  19$ \pm $3	    \\
  6802    &   500     &  15$ \pm $2	    \\
  7302    &   500     &  16.6$ \pm $1.8     \\
  7802    &   500     &  13.0$ \pm $2.0     \\
  8302    &   500     &  14.8$ \pm $1.7     \\
  13302   &   500     &   9.7$ \pm $1.5     \\
  13802   &   500     &  10.8$ \pm $1.5     \\
  14302   &   500     &  8.8$ \pm $1.4     \\
  19202   &   900     &  6.1$ \pm $1.2     \\
  20102   &   900     &  6.7$ \pm $1.0     \\
  25052   &   1000    &  6.0$ \pm $1.2     \\
  26052   &   1000    &  5.0$ \pm $1.0     \\
  30482   &   2660    &  3.5$ \pm $0.7     \\
  40002   &   10700   &  2.7$ \pm $0.6     \\
  50702   &   10700   &  3.1$ \pm $0.6     \\
  61402   &   10700   &  2.7 $ \pm $0.5    \\
  281121  &   52356   &  0.53$ \pm $0.12   \\
  418938  &   80974   &  0.47$ \pm $0.13   \\
  618410  &   100110  &  0.24$ \pm $0.05   \\
  1492073 &   353910  &  0.044$ \pm $0.010   \\
  1842640 &   173812  &  0.050$ \pm $0.010   \\
  2476438 &   225974  &  0.035$ \pm $0.012 \\\hline
\end{tabular}
\end{table}

\section{XRT data analysis}
\label{xrt}
\indent

XRT observations of the GRB\,050408 field started on 2005 April 8 at
17:05:17 UT (2545~s after the HETE-II trigger); 
data collection in photon counting (PC) mode started a few seconds later 
(see \citet{hill04} and \citet{modes2} for a description of XRT readout modes).
\textit{Swift} subsequently continued monitoring 
the GRB field at later times, collecting a total of 12 observations over a period of 38~d.
The total exposure time in PC mode was $\sim 223$~ks.
Occasionally, due to the high background, the XRT 
switched into windowed timing mode.
However, the source count rate was very low in these frames, 
and therefore these data were not included in our analysis.

The XRT data from the SDC archive at the NASA Goddard Space Flight Center were processed with the
{\tt XRTDAS} \footnote{\scriptsize{\texttt{http://swift.gsfc.nasa.gov/docs/swift/analysis/xrt\_swguide\_v1\_2.pdf}}} software package (v. 1.7.1) developed at the ASI Science
Data Center. Calibrated and cleaned level-2 event files were produced
with the {\tt xrtpipeline} task, applying the standard screening criteria:
frames with a CCD temperature 
greater than $-47$~$^\circ$C were rejected, and
bad pixels and bad aspect time intervals were eliminated.

   \begin{figure}
   \centering
   \includegraphics[width=\columnwidth]{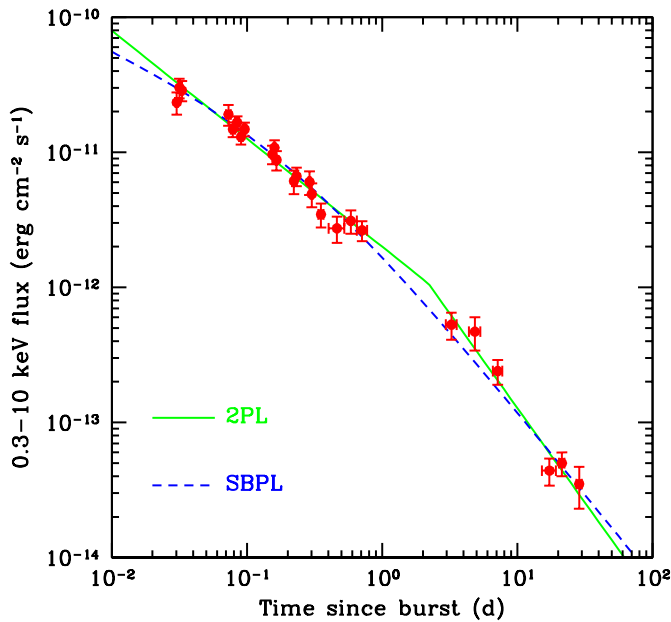}

      \caption{The 0.3--10 keV X-ray light curve of the afterglow of
      GRB\,050408 (PC mode data only). The solid and dashed lines show
      the best-fit models assuming a sharply- (2PL) and smoothly- (SBPL)
      broken power law, respectively.\label{fg:lc}}

   \end{figure}

\begin{table*}
\caption{Best-fit parameters with their associated 68\% confidence
intervals, for different functional forms: single power law (PL),
sharply broken power law (2PL), smoothly broken power law (SBPL), and
Beuermann fit \citep[B;][]{Beuer99} for two values of the smoothness
parameter $s$.\label{tb:lc}}
\begin{tabular}{ll|ll|ll|ll}\hline\hline
Model & $\chi^2_{\rm r}$ (dof) & \multicolumn{2}{c|}{$\alpha_1$}  & \multicolumn{2}{c|}{$\alpha_2$} &\multicolumn{2}{c}{$t_{\rm b}$ (d)} \\
      &          & Best  & 68\% C.I.                       & Best & 68\% C.I.               & Best & 68\% C.I.\\ \hline
PL    & 2.50 (23)     & 0.99  & $0.97<\alpha_1<1.01$            & ---  & ---                     & ---  & --- \\
2PL   & 1.0 (21)     & 0.80  & $0.49<\alpha_1<0.67, 0.76<\alpha_1<0.84$ & 1.40 & $1.25<\alpha_2<1.55$           & 2.2 & $0.12<t_{\rm b}<0.18, 1.3<t_{\rm b}<3.0$ \\
SBPL  & 0.92 (21)     & 0.46  & $-0.12<\alpha_1<0.78$                  & 1.24 & $1.11<\alpha_2<1.91$           & $0.21$ & $0.05<t_{\rm b}<6.6$ \\
B, $s=0.1$& 0.89 (21) &0.21   & $\alpha_1 < 0.31$  &2.05  & $1.81<\alpha_2<2.89$  & \multicolumn{2}{l}{unconstrained} \\
B, $s=0.5$& 0.89 (21) &0.13   & $\alpha_1 < 0.75$ &1.51  & $1.15<\alpha_2<3.10$ & \multicolumn{2}{l}{unconstrained} \\\hline
\end{tabular}
\end{table*}

\subsection{Image analysis}
\indent

The 0.3--10 keV PC mode image of the field was analyzed with the {\tt XIMAGE} package
(v.{} 4.3). A previously uncatalogued, bright X-ray source was clearly detected
inside the HETE-II error circle, at the coordinates
RA(J2000)$\mbox{} = 12^{\mathrm h}02^{\mathrm m}17\fs29$,
Dec(J2000)$\mbox{} = 10\degr51\arcmin11\farcs4$ (3\farcs5 error radius; 90\%
containment). 
This position and its error were evaluated taking into account the
latest XRT boresight calibration \citep{Moretti05}.
The XRT coordinates are 60\arcsec{} and 1\farcs95 away from the HETE-II
error box center and the optical afterglow position, respectively.
Another non-fading source with count rate $(2.9\pm0.1)\times 10^{-3}$
count~s$^{-1}$ is present inside the HETE-II error circle, at a distance
of $\approx 38\arcsec$ from the afterglow.  To avoid contamination from
this source, the events for the temporal and spectral analysis were
selected from
a circular region of 8 pixels (19\arcsec) radius centered at the
afterglow position, which contains about 77\% of the photons at 1.5 keV
\citep{moretti}. The appropriate ancillary response file was used to correct
for the PSF losses.

\subsection{Temporal analysis}
\indent

For the temporal analysis, the standard grade selection for PC mode
(grades from 0 to 12) was adopted, in order to maximize the
statistics. Only the 0.3--10 keV energy range was considered.
The background in the extraction region was evaluated by 
comparing several source-free boxes 
in the field and, since it was stable during the whole observation,
a constant level of $1.3 \times 10^{-4}$ count~s$^{-1}$ was subtracted from the light curve.
Moreover, the contribution of the nearby serendipitous source was evaluated computing the
number of photons falling inside the afterglow extraction region 
and the afterglow count rate was corrected for this contamination.
The light curve was binned to ensure a minimum of 20 counts per bin.
The data 
are reported in Table ~\ref{tb:lcdata}. 
The count rate was converted to an unabsorbed 0.3--10 keV flux using a
conversion factor of $1.0\times10^{-10}$ erg cm$^{-2}$ count$^{-1}$,
obtained using the results from the spectral analysis (Sect~2.3).
The light curve is shown in Fig.~\ref{fg:lc} (points).
A gap of approximately two days is present in the data due to the close
proximity of the Moon preventing the NFIs from observing.

\begin{figure}
\includegraphics[width=\columnwidth]{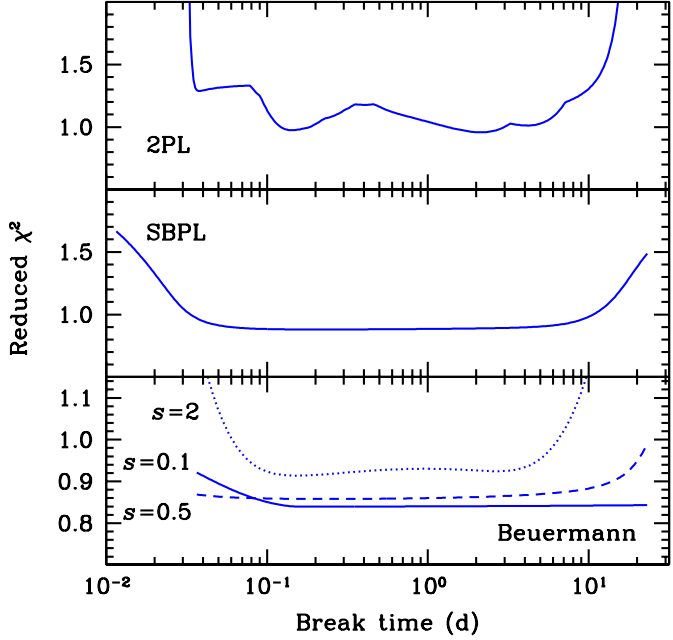}
\caption{Reduced $\chi^2$ as a function of the break time $t_{\rm b}$
for three functional forms: from top to bottom, joint broken power law
(2PL), smoothly broken power law (SBPL), and the Beuermann fit, for
different values of the smoothness parameter $s$
\citep{Beuer99}.\label{fg:chisq}}
\end{figure}

We initially fit the afterglow decay with a single power law, obtaining an
unacceptable $\chi^2_{\rm r} = 2.5$ (23 degrees of freedom, d.o.f.).
Looking at the residuals, a systematic deviation of the data from the model was
apparent, showing a continuous steepening of the light curve with time.
We thus fitted the data with a broken power law ($F \propto t^{-\alpha_1}$ for
$t < t_{\rm b}$, $F \propto t^{-\alpha_2}$ for $t > t_{\rm b}$), where
$\alpha_1$ and $\alpha_2$ are the early- and late-time decay slopes,
respectively, and $t_{\rm b}$ is the break time. The best fit
(Fig.~\ref{fg:lc}, solid line) significantly improved, yielding $\chi^2_{\rm r}
= 1.0$ for 21 d.o.f. (null hypothesis probability of $6.6 \times 10^{-5}$,
according to an $F$-test). The best fit parameters, together with their
confidence intervals, are reported in Table~\ref{tb:lc}. Errors were computed
leaving all parameters free to vary. The fit, and in particular $t_{\rm b}$, is
not well constrained. Fig.~\ref{fg:chisq} (upper panel) shows the behaviour of
$\chi^2_{\rm r}$ as a function of $t_{\rm b}$. Two minima are apparent at
approximately the same level (with the larger value of $t_{\rm b}$ being
slightly preferred). In an independent analysis, \citet{Foley05} indicated the
lower value as the best fit. The reason for the discrepancy is possibly due to
minor differences in the reduction process (e.g. background subtraction, data
binning). However, our results are 
 consistent with those derived by
\citet{Chinca05} and \citet{Nousek05} (who analyzed only the first part of the
light curve). The smoothness of the light curve prompted us to attempt
different functional forms, using the so-called smoothly-broken power law
(SBPL) model: $F(t) \propto 1/[(t/t_{\rm b})^{\alpha_1} + (t/t_{\rm
b})^{\alpha_2}]$. This provided a slightly better fit ($\chi^2_{\rm r} = 0.92$
for 21 d.o.f.; dashed line in Fig.~\ref{fg:lc}), but $t_{\rm b}$ could not be
constrained (Fig.~\ref{fg:chisq}, middle panel). 
Similarly,  $t_{\rm b}$ could not be
constrained with a fit to the Beuermann model ($F(t) \propto [(t/t_{\rm
b})^{s\alpha_1} + (t/t_{\rm b})^{s\alpha_2}]^{-1/s}$; \citealt{Beuer99}). The
parameter $s$ (the smoothness index)  controls how fast the transition between
the two phases happens:  large values of $s$ imply a sharp break. Leaving all
parameters free, 
small  values of $s$ (i.e., a smooth break) were  systematically
preferred ($s = 0.1$--0.5) but the fit did not converge.  Fixing $s$ to a set
of definite values allowed 
the confidence intervals  for the parameters to be determined
(Fig.~\ref{fg:chisq}, lower panel).

Finally, in consideration of the fact that the values of the last three data points of the light curve 
are consistent within the errors, we checked for the presence of a persistent X-ray emission 
coming from the afterglow host. 
We found that adding a constant component to the broken power law model (2PL) 
does not yield a significant improvement to the fit ($F$-test probability of 
a chance improvement $\sim 80\%$).

\begin{table*}
\caption{Results of the spectral fits to the X-ray afterglow of
GRB\,050408. An asterisk indicates a frozen parameter. The redshifts for
the Galactic and host absorption were kept fixed at $z = 0$ and $z =
1.2357$, respectively. The errors are at 90\% confidence level for one
interesting parameter.\label{bestfit}}
\centering
\begin{tabular}{l|ccc|cccc} \hline\hline
             &\multicolumn{3}{c|}{\rm Galactic absorption}                        &\multicolumn{4}{c}{\rm Galactic + host absorption}                                    \\ \hline
Time range   &$\beta$             &$N_{\rm H}$         &$\chi^2_{\rm r}$ (d.o.f.) &$\beta$             &Galactic $N_{\rm H}$ &Host $N_{\rm H}$    &$\chi^2_{\rm r}$ (d.o.f.) \\
             &                    &$10^{21}$~cm$^{-2}$ &                          &                    &$10^{21}$~cm$^{-2}$  &$10^{21}$~cm$^{-2}$ &                      \\ \hline
All data     &$1.3\pm0.2$        &$2.9\pm0.6$          &1.4 (39)                  &$1.1\pm0.1$         &$0.174^*$            &$12^{+4}_{-3}$      &1.5 (39)                  \\
First part   &$1.3\pm0.2$         &$3.0\pm0.7$         &1.2 (32)                  &$1.2\pm0.2$         &$0.174^*$            &$15\pm4$            &1.3 (32)                  \\
Second part   &$1.2^{+0.4}_{-0.3}$ &$3.0^*$             &1.1 (6)                   &$1.0\pm0.3$         &$0.174^*$            &$15^*$              &0.9 (6)                   \\ \hline
\end{tabular}
\end{table*}

   \begin{figure}
   \centering
   \includegraphics[angle=-90,width=\columnwidth]{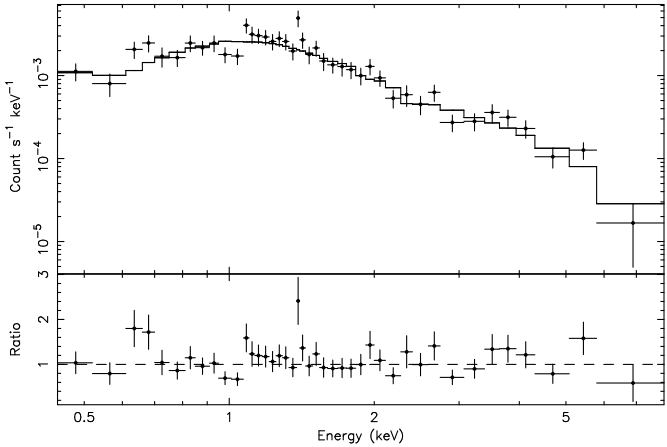}
      \caption{The average 0.3--10 keV X-ray spectrum of the GRB\,050408
      afterglow (top panel), together with the best-fit absorbed power-law
      model (with Galactic and host absorption). The bottom panel shows
      the ratio between the data and the best-fit model.\label{sp}}
   \end{figure}

\subsection{Spectral analysis}
\indent

Events were extracted for the spectral analysis from the same circular
region used for the 
temporal analysis, but a more strict selection on
event grades was applied (grades 0-4, i.e. single- and double-pixel
events), in order to achieve better spectral resolution.  The spectrum
was binned to ensure a minimum of 20 counts per energy bin. Channels
below 0.3 keV and above 10.0 keV were excluded.
We first produced an average spectrum obtained by summing all the available
observations.
A spectral fit was done using the {\tt XSPEC} package (v. 11.3.1).
We modelled the spectrum with an absorbed power law of spectral index $\beta$ and
Hydrogen column density $N_{\rm H}$ (assumed to be at redshift $z = 0$  and with 
the heavier elements fixed at solar abundances).
The best-fit parameters are shown in Table~\ref{bestfit}.
The value of $N_{\rm H}$ was found to be significantly higher than the
Galactic 
value ($N_\mathrm{H} = 1.74\times10^{20}~\mathrm{cm}^{-2}$,
\citealt{dickey}; $N_\mathrm{H} = 1.54\times10^{20}~\mathrm{cm}^{-2}$, \citealt{Kalberla05}).
We therefore included an 
additional absorption component at the redshift of
GRB\,050408 (model {\tt ZWABS} in {\tt XSPEC}), keeping
values of the redshift ($z = 1.2357)$ and the Galactic column density
($N_\mathrm{H} = 1.74\times10^{20}~\mathrm{cm}^{-2}$ at $z = 0$) frozen.
The fit provided $N_{\rm H} = 1.2_{-0.3}^{+0.4}\times 10^{22}$~cm$^{-2}$ for
the redshifted absorber. The 0.3--10 keV XRT
spectrum is shown in Fig.~\ref{sp}, together with the best fit absorbed power
law model. Our results are consistent with the analysis reported by
\citet{Foley05}. The average unabsorbed 0.3--10 keV flux, calculated between
$\sim 2.5$~ks and $\sim 67$~ks after the trigger, is $7.1\times10^{-12}$ erg
cm$^{-2}$ s$^{-1}$.

We looked for spectral variability across the observation. 
Due to the limited statistics at late times, we 
split the data into two bins, performing a power law fit (with absorption
both in the Milky Way and in the host galaxy) to both sections.
The best-fit parameters were consistent with those found for
the average spectrum (Table~\ref{bestfit}).
For the second part of the light curve, we had to freeze the column
densities (which are not expected to change) to the values measured
before the break. As can be seen from Table~\ref{bestfit}, the results
from the two phases are consistent.
To further test this result, we studied the time evolution of the
hardness ratio, computed as the ratio between the count rate in the
2--10 and 0.3--2 keV bands. No significant variability was apparent.

\section{Discussion}
\label{discussion}
\indent

The \textit{Swift} observations allowed a detailed study of the X-ray light curve of
GRB\,050408, extending for more than three decades in time. This in turn led to
the detection of a very smooth break in the temporal decline, which would have been 
more difficult to detect with 
less coverage. The initial portion of the light curve presents a 
shallow decay, followed by a steepening to a more conventional slope. This is consistent with the typical
behaviour observed in several \textit{Swift} GRBs \citep{Chinca05,Nousek05,OBrien05},
assuming that the early rapid decline may have occurred during the 42
min between the trigger and the first observation.
The spectral slope is also typical of GRB afterglows \citep{DePasq05}.

What is less typical is the smoothness of the transition between the flat and
steep phases. Despite 
the long-duration 
coverage, there is no clear evidence of a well-defined break time, $t_{\rm b}$. 
Indeed, fits to the light curve allow a
very broad range for $t_{\rm b}$. Moreover, fits with a smooth transition are
preferred. The behaviour has been observed in very few afterglows. 
Most optical light curves, when fitted with the Beuermann model, have smoothness parameters as
large as $s = 10$ \citep{Covino03,Zeh05}. In the X-ray bandpass, 
the statistics are usually worse, but joint power laws usually fit 
the data well. Without the
long-term coverage provided by \textit{Swift}, the curvature of this break
would have easily been missed.

The early part of the light curve ($t < t_{\rm b}$) is too shallow to be
explained in terms of the standard afterglow models
\citep[e.g.][]{Sari98,PanKum00}, either considering a homogeneous (ISM)
or a wind-shaped surrounding medium. The early-time slope
is $\alpha_1 < 0.84$ for all models (see
Table~\ref{tb:lc}). Given the spectral slope%
\footnote{We denote with $\beta_1$ and $\beta_2$ the spectral slopes
before and after the break, respectively.} $\beta_1 = 1.2 \pm 0.2$,
we have $\alpha_1 - 3\beta_1/2 < -0.96$, which is inconsistent with the
expected values: $0$ (ISM, $\nu < \nu_{\rm c}$), $+1/2$ (wind, $\nu <
\nu_{\rm c}$), or $-1/2$ ($\nu > \nu_{\rm c}$), where $\nu_{\rm c}$ is
the cooling frequency. Fast-cooling models are also ruled out, since a spectral index of
$\beta_1 = 0.5$ would be expected.

There are several ways to explain a shallow slope. For example, if the
cooling of electrons is dominated by inverse Compton rather than
synchrotron losses (that is, the Compton parameter is $Y \gg 1$), the
decay above $\nu_{\rm c}$ is expected to be flatter than in the standard
case \citep[e.g.][]{SariEsin01,PanKum00}. In fact, the cooling frequency
is $\nu_{\rm c} = \nu_{\rm c}^{\rm s} (1+Y)^{-2}$ (where $\nu_{\rm
c}^{\rm s}$ is the synchrotron cooling frequency), and it is easy to
show that the flux above $\nu_{\rm c}$ is $F_\nu(t) = F_\nu^{\rm s}(t)
(1+Y)^{-1}$, where $F_\nu^{\rm s}$ is the synchrotron flux in the case
of no Compton losses. Since the Compton parameter decreases with time
(at least in the slow cooling regime), when $Y \gg 1$ the decay law is
flatter than in the case of pure synchrotron. The steepening of the
light curve pinpoints the time at which $Y = 1$ (so that $F_\nu \sim
F_\nu^{\rm s}$). The smoothness of the transition would imply that the
decrease of $Y$ with time is very slow, possibly not following a power
law. Given the soft X-ray spectral index, the Compton component should
be confined to energies above the XRT range.

Observations in the optical show that this band likely lies below the cooling
frequency \citep{Foley05,tng}. The light curve at these frequencies may
have a decay similar to that in the X-ray band, thus disfavoring this
assumption. However, the poorly constrained X-ray decay and the presence
of significant extinction in the optical makes more stringent
comparisons difficult.

An alternative way to explain the early flat decay is to have an outflow
with an angular structure \citep{Rossi02,ZhangMesz02}, that is a jet
with an energetic core and less intense wings. As time elapses, the
fireball Lorentz factor, $\Gamma$, decreases so that a larger portion of
the jet becomes visible to the observer due to relativistic
aberration. If the jet is viewed off-axis (at an angle $\vartheta_{\rm
v}$), the core contribution becomes more and more important, so that its
increased emission partly compensates for the flux decline, leading to a
slower decay. In this case, the light curve break would happen when
$\Gamma \sim 1/\vartheta_{\rm v}$. \citet{PanKum03} modeled the observed
emission from off-axis structured jets, showing that flat ($\alpha_1
\approx 0.5$) early-time slopes are possible for $\vartheta_{\rm
v}/\vartheta_{\rm c} \sim \mbox{a few}$, where $\vartheta_{\rm c}$ is
the angular extension of the jet core. In this case, however, a sharp
break is predicted \citep[$s > 1$;][]{Rossi04}. Thus, the smoothness of the
observed transition may 
indicate that the energy angular profile of
the jet was not a pure power law as function of the off-axis angle.

The final possibility is that the early shallow decay is due to delayed energy
injection in the fireball \citep[e.g.][]{DaiLu98,Panaitescu98,ZhangMesz01}.
This can be achieved in two ways: long-lasting energy emission from the central
engine, or refreshed shocks from slow shells catching the leading fireball
after it has decelerated \citep[see e.g.][for a general overview]{Zhang05}.
Even if the latter model may be preferred on theoretical grounds (no extended
activity is required), the two scenarios are difficult to distinguish
observationally. Both can account for light curves as flat as $F(t) \propto
t^0$. The smoothness of the break may indicate that, in this case, the additional energy injection
did not follow a power law in time, but had a more complex
history. 

However, an effective index $q$ can be introduced by parametrizing the
energy injection as $\dot{E} \propto t^{-q}$. A viable solution is found with
$q < 0.70$ (for $\alpha_1 < 0.84$), assuming $\nu > \nu_{\rm c}$ (either ISM or
wind). The case with $\nu_{\rm i} < \nu < \nu_{\rm c}$ ($\nu_{\rm i}$ being the
synchrotron injection frequency) is acceptable only in the ISM case (providing
$q < 0.48$), while the wind case is excluded since $q < 0$ would be required.
Moreover, if $\nu < \nu_{\rm c}$, a very large 
$p = 3.2 \pm 0.2$ would be
implied, so that we favor the case $\nu > \nu_{\rm c}$. Values of the order $q
\approx 0.5$ have been inferred also for other GRB afterglows
\citep{Zhang05,Nousek05}. If energy was supplied to the fireball through slow
shells impacting the decelerated fireball, and parametrizing their Lorentz factor
distribution as $M(>\gamma) \propto \gamma^{-r}$ ($M(\gamma)$ 
being the mass of shells with Lorentz factor larger than $\gamma$),
the above values for $q$
correspond to $r > 1.9$ (ISM, $\nu > \nu_{\rm c}$), $r > 2.7$ (wind, $\nu >
\nu_{\rm c}$), and $r > 2.7$ (ISM, $\nu_{\rm i} < \nu < \nu_{\rm c}$). In the
context of the injection model, two possibilities are viable to explain the
observed break in the light curve. First, $t_{\rm b}$ may simply identify the
end of the energy injection process. The decay after $t_{\rm b}$ would then
correspond to a standard isotropic afterglow. Indeed, the observed $\alpha_2
\approx 1.5$ and $\beta_2 \approx 1.0$ satisfy, within the errors, several
closure relations (again, only the wind case with $\nu_{\rm i} < \nu < \nu_{\rm
c}$ is unfavored). An alternative explanation is that the injection did not
stop at $t_{\rm b}$, and the steepening was due to a jet effect
\citep{Rhoads99,Sari99}. 
The data (Table~\ref{tb:lc}) are consistent with a
late-time decay $\alpha_2 < 2$ (particularly if the break time was early),
which cannot be accounted for in the standard model, where $\alpha_2 = p \ga 2$
is predicted. Energy injection 
continuing after the jet break would naturally
make the decay shallower. Indeed, there are examples where energy injection
lasted for considerable durations 
 \citep[e.g. XRF\,050406, where this phase lasted
for $>10^6$~s;][]{Romano05}. A detailed analysis of the dynamics of a
relativistic jet being refreshed is beyond the 
scope of this work, however we
envisage this solution as qualitatively possible. Finally we 
note that, within the
uncertainties, our data do not exclude steeper slopes ($\alpha_2 > 2$), so that
energy injection is not strictly required after the break. However, it would be
coincidental that the cessation of the refreshing and the jet break 
happened at the same time.

The long-lasting injection episode implies that the fireball energy
increased significantly. 
 \citet{Nousek05} give several recipes to estimate the fractional
energy increase, $f$. Limiting ourselves to the regime $\nu > \nu_{\rm
c}$, and conservatively only taking the evolution where 
 $t < t_{\rm b}$ into account, we have $f = (t_{\rm b}/t_{\rm start})^\kappa$, where
$t_{\rm start} < 2500$~s is the start of the injection phase, $\kappa =
2\Delta\alpha/(1+\beta) \approx 0.5$ and $\Delta\alpha$ is the
difference between the observed decay slope and the one which we would
see in the case of no injection ($\alpha = 3\beta/2-1/2$). With these
numbers, and taking $t_{\rm b} \sim 10^5$~s, we find $f > 2 \div 25$, so
that the fractional energy increase 
was quite significant. In the case where $\nu <
\nu_{\rm c}$ (which we do not favor), $f$ would be even larger.

There is another remarkable feature concerning GRB\,050408, namely its large
rest-frame column density, of the order of $10^{22}$~cm$^{-2}$  (computed
assuming solar abundances). Such a value is typical of giant molecular clouds
\citep{ReichartPrice02}. Based on \textit{Beppo}SAX and XMM-\textit{Newton}
data, many authors have reported column densities as large as
$10^{22}$~cm$^{-2}$ \citep[e.g.][]{GalamaWijers01,Watson02,Stratta04,DeLuca05}.
This was later confirmed by \citet{Campana05} using \textit{Swift}-XRT data,
showing that excess column density is a common feature among GRB afterglows.
So, it is not surprising to find such a value. It is however more difficult to
explain the relative brightness of the optical afterglow. Using the Galactic
gas-to-dust ratio \citep{PredehlSchmitt95}, the measured column density would
correspond to $A_V \sim 8$~mag. This is not consistent with the detection of an
optical counterpart (the observed $R$ band would suffer 14~mag of extinction at
$z = 1.2357$). \citet{Foley05}, from optical spectroscopy and photometry,
estimated $A_V \sim 0.5$--1~mag, assuming a SMC-like extinction curve.
The small amount of optical extinction compared to the X-ray absorption was
already noticed in previous cases \citep{GalamaWijers01,Stratta04}, and
ratios $A_V/N_{\rm H}$ as low as ten times less than the Milky Way value have been
reported. This implies either different dust optical properties or a low
dust-to-metals ratio.

In the former case, the estimation of the dust content from the afterglow
spectral properties may be incorrect. For example, from the analysis of the
absorbing element abundances in the GRB\,020813 afterglow,
\citet{SavaglioFall04} derived an amount of dust larger than inferred from the
analysis of the continuum spectral shape, thus indicating an anomalous
extinction curve. However, \citet{Foley05} detected Titanium overabundance in
the optical spectrum of the GRB\,050408 afterglow. Since this element is highly
refractory, low dust content (or a different dust composition) is inferred for
this line of sight, independently on its transmission properties. This would
therefore leave us with the second possibility, namely an intrinsically low
dust-to-metals ratio. Such composition may be a property of the GRB formation
environments, where the intense UV radiation field from the young, hot stars
may hamper dust formation. Also, the young stellar age of GRB hosts may imply
that dust has not yet had time to form \citep{Watson06}. Finally, a low dust
content may be the direct effect of the burst explosion, which is able to
sublimate dust grains up to $\sim 100$~pc from the explosion site
\citep{WaxmanDraine00,Fruchter01}.

\section{Conclusions}
\label{conclusions}
\indent

GRB\,050408, detected by HETE-II, was monitored by \textit{Swift}-XRT over a
period of 38~d. This allowed a detailed study of the X-ray light curve,
extending over more than three decades in time. A very smooth break was
apparent $\sim 2 \times 10^5$~s after the trigger, separating a shallow decay
phase from a steeper one. 
The transition was extremely gradual, with no definite break time. 
This is in contrast to 
the usual behaviour observed for X-ray
and optical afterglows, where a sharp and abrupt break is 
usually observed. On the other hand, the spectral properties 
of this afterglow were not unusual, and
no spectral variability was observed. The X-ray spectrum showed a large Hydrogen
column density ($N_\mathrm{H} \sim 10^{22}$~cm$^{-2}$), significantly in excess 
of the Galactic value.
This may be a common feature among GRB
afterglows \citep{Campana05,GalamaWijers01,Stratta04}, possibly indicating
dense environments \citep{ReichartPrice02}. This large X-ray absorption was,
however, accompanied by a relatively small optical extinction
\citep{Foley05,tng}.

The first portion of the afterglow light curve is too shallow to be
explained in terms of standard afterglow models. Alternative solutions
were considered, including the possibility of a structured jet observed
off-axis, or the presence of Compton radiation affecting the decay above
the cooling frequency. Our data are also consistent with late injection of
energy from the central engine lasting for considerably longer than 
the GRB explosion. In this case, the break may pinpoint the end of the
injection activity. Alternatively, the steepening may be a standard jet
break. In this case the injection has to continue even during the steep
phase.
In both cases, the energy supplied by the central engine is
considerable, being larger by a factor of $> 2 \div 25$ 
than the initial amount. This may pose serious efficiency problems to GRB radiation
models.

\begin{acknowledgements}
We thank Don Lamb and Nat Butler for useful discussions, Ryan Foley
and Dave Pooley for providing us their X-ray light curve. We thank
F. Tamburelli and B. Saija for their work on the XRT data reduction
software.  We acknowledge financial support by the Italian Space Agency
(ASI) through grant I/R/039/04 and through funding of the ASI Science
Data Center. DM and OG gratefully acknowledge MIUR and PPARC funding.
\end{acknowledgements}

\end{document}